\documentclass[onecolumn,showpacs,showkeys,amsmath,amssymb,superscriptaddress,floatfix,nofootinbib]{revtex4}

\usepackage{graphicx}
\usepackage{bm} 
\usepackage{subfigure}
\usepackage[colorlinks=true,linktocpage=true,linkcolor=blue,citecolor=blue]{hyperref}
\usepackage[usenames,dvipsnames]{color}

\usepackage{setspace,ulem,color}
\usepackage{amsfonts}
\usepackage{amssymb}
\usepackage{amsmath}
\usepackage{graphicx}

\newcommand{\coll}{{\cal C}}
\newcommand{\deltaf}{\delta\! f}
\newcommand{\tdeltaf}{{\tilde \deltaf}}
\newcommand{\ped}{{\cal E}}
\newcommand{\pres}{{\cal P}}
\newcommand{\preseq}{\pres_{\rm eq.}}
\newcommand{\taueq}{\tau_{\rm eq.}}
\newcommand{\feq}{f_{\rm eq.}}

\begin{document}

\title{
Anisotropic matching principle for the hydrodynamics expansion
}

\author{Leonardo Tinti} 
\email{dr.leonardo.tinti@gmail.com}
\affiliation{Institute of Physics, Jan Kochanowski University, PL-25406~Kielce, Poland}

\begin{abstract}

Following the recent success of anisotropic hydrodynamics we propose a new, general prescription for the hydrodynamics expansion around an anisotropic background. The anisotropic distribution is fixing exactly the complete energy-momentum tensor, just like the effective temperature is fixing the proper energy density in the ordinary expansion around local equilibrium. This means that momentum anisotropies are already included at the leading order, allowing for large pressure anisotropies without the need of a next to leading order treatment. The first moment of the Boltzmann equation (local four-momentum conservation) provides the time evolution of the proper energy density and the four velocity. Differently from previous prescriptions, the dynamic equations for the pressure corrections are not derived from the zeroth or second moment of the Boltzmann equation, but they are taken directly from the exact evolution given by the Boltzmann equation. We check the effectiveness of this new approach by matching with the exact solution of the Boltzmann equation in the Bjorken limit with the collisional kernel treated in relaxation time approximation, finding an unprecedented agreement.

\end{abstract}



\maketitle

%%%%%%%%%%%%%%%%%%%%%%%%%%%%%%%%%%%%%%%%%%%%%%%%%%%%%%%%%%%%%%%%%%%%%%%%%%%%%%%%%%%%%%%%%%%%%%%%%%%%
\section{Introduction}
\label{sect:intro}
%%%%%%%%%%%%%%%%%%%%%%%%%%%%%%%%%%%%%%%%%%%%%%%%%%%%%%%%%%%%%%%%%%%%%%%%%%%%%%%%%%%%%%%%%%%%%%%%%%%%%

Relativistic viscous hydrodynamics  has been an important tool for understanding the evolution of matter in the extreme conditions of heavy-ion collisions at RHIC (Relativistic Heavy-Ion Collider) and the LHC (Large Hadron Collider). Therefore a large interest has been triggered in the development of the hydrodynamic framework
\cite{Israel:1976tn,Israel:1979wp,
Muronga:2001zk,Muronga:2003ta,
Heinz:2005bw,
Baier:2006um,Baier:2007ix,
York,Teanay,
Romatschke:2007mq,Dusling:2007gi,Luzum:2008cw,
Song:2008hj,El:2009vj,PeraltaRamos:2010je,
Denicol:2010tr,Denicol:2010xn,
Schenke:2010rr,Schenke:2011tv,
Bozek:2009dw,Bozek:2011wa,
Niemi:2011ix,Niemi:2012ry,
Bozek:2012qs,Denicol:2012cn,Jaiswal:2013npa,
PeraltaRamos:2012xk}. 
Despite the succesfull use of viscous hydrodynamics, some theoretical shortcoming have to be properly addressed. Several studies pointed out that viscous corrections combined with rapid longitudinal expansion (caused by large longitudinal gradients) result in a substantial pressure asymmetry at early times, even at a very low viscosity to entropy density ratio, $4\pi \eta/{\cal S} \sim 1$.  Similarly, a variety of microscopic models (string models, color glass condensate, pQCD kinetic calculations) predict large momentum anisotropies at early times. On the other hand, in the limit of infinitely strong coupling, where the AdS/CFT correspondence can be used as an effective model, one finds a similar situation --- a large difference between the pressures in the transverse plane and in the longitudinal direction \cite{Heller:2011ju,Heller:2012je}, persisting for a significant fraction of the evolution.  Large pressure anisotropies are a cause for concern, since viscous hydrodynamics treats pressure corrections in a perturbative manner and, indeed, the presence of very large shear corrections (of the order of the isotropic pressure) may lead to unphysical results such as the negative longitudinal pressure and/or negative one-particle distribution functions. 

A new approach to treat these problems is {\it anisotropic hydrodynamics} (aHydro) \cite{Florkowski:2010cf,Martinez:2010sc,
Ryblewski:2010bs,Martinez:2010sd,
Ryblewski:2011aq,Martinez:2012tu,
Ryblewski:2012rr,Ryblewski:2013jsa,
Florkowski:2012ax,Florkowski:2012as,
Florkowski:2014txa,
Florkowski:2014sfa,
Florkowski:2011jg}. The main feature of this approach, in contrast with the conventional approach, is to treat the expected high degree of pressure anisotropy of the produced matter already in the leading order of the hydrodynamic expansion. In the original formulation of aHydro, the leading order includes a single anisotropy parameter taking into account a large difference between the longitudinal and transverse pressures. This approach is unable to reproduce the pressure anisotropy in the transverse plane, which is generated by the radial flow \cite{Florkowski:2011jg}. However, non-trivial transverse dynamics is essential for the correct description of transverse collective behavior, like the elliptic flow, $v_2$, suppression. 

In Ref.~\cite{Bazow:2013ifa},  Bazow, Heinz, and Strickland extended the aHydro framework to the (2+1)-dimensional case, which is longitudinally boost invariant but non-homogeneous in the transverse plane. This approach has been recently extended to massive (non-conformal) systems in Ref.~\cite{Bazow:2015cha}. In Refs.~\cite{Bazow:2013ifa,Bazow:2015cha}, however, the transverse pressure asymmetries (and part of the longitudinal pressure correction) are included only through next to leading order corrections. On the other hand, it has been demonstrated (see Ref.~\cite{Tinti:2014conf,Florkowski:2014nc,Nopoush:2014nc,Nopoush:2015gf}) that it is possible to generalize the background to the (1+1)-dimensional case (boost invariant and cylindrically symmetric radial flow). Very recently, we have proposed a set of equations for anisotropic hydrodynamics that is consistent with the full (3+1)-dimensional expansion~\cite{Tinti:2014yya}.

In all of the previous works on anisotropic hydrodynamics there was an ambiguity in the choice of the dynamical equations of motion, both for the leading order treatment and the next to leading order of the hydrodynamics expansion. Besides the local four-momentum conservation (first moment of the Boltzmann equation), at least some of the remaining equations have been derived from the zeroth moment (typically describing particle creation), the second moment of the Boltzmann equation or both. However, in principle any other moment of the Boltzmann equation could provide the necessary equations. This is not a new problem, since in the ordinary viscous expansion there is exactly the same ambuguity~\cite{Denicol:2014loa}. As pointed out in Ref.~\cite{Denicol:2014loa}, this problem has already been solved for the hydrodynamics expansion around local equilibrium. The Boltzmann equation can be used to provide directly the evolution equation for the moments of the Boltzmann {\it distribution} (not to confound with the moments of the Botzmann {\it equation}). Among them, for instance, there are the shear pressure corrections $\pi^{\mu\nu}$, and the higher moments which appear in the exact evolution of $\pi^{\mu\nu}$ itself. This approach has been very successful for second order viscous hydrodynamics~\cite{Denicol:2014sbc,Jaiswal:2014}, however, it can not be extended directly to the anisotropic expansion. Since the anisotropic background has at least one extra free degree of freedom (anisotropy parameter, contributing to the shear viscous corrections), the exact evolution from the Boltzmann equation (for instance the evolution of $\pi^{\mu\nu}$) will not be able to fix both the leading order and the next to leading order contribution (in previous works denoted as $\pi^{\mu\nu} = \pi^{\mu\nu}_a+\tilde{\pi}^{\mu\nu}$, with $\pi^{\mu\nu}_a$ depending only on the anisotropic background~\cite{Bazow:2013ifa,Bazow:2015cha,Tinti:2014yya}). 

In this paper we propose to use a generalized version of the Landau matching in order to determine extra degrees of freedom related to anisotropy. Just like the effective temperature fixes the proper energy density at all orders in the viscous hydrodynamics expansion, we propose that the extra parameters in the anisotropic distribution will fix the pressure corrections in the anisotropic expansion (in this case $\pi^{\mu\nu}\equiv\pi^{\mu\nu}_a$ at all orders by definition). In order to illustrate our new method, we consider a particular application of the proposed prescription, namely, we deal with the leading order of the expansion with the collisional kernel treated in the relaxation-time approximation, and we test it against the exact solutions in the Bjorken flow limit. The agreement is significantly improved compared both to the best prescription for leading order anisotropic hydrodynamics, and to the best prescription for second order viscous hydrodynamics (which is proved to be the close-to-equilibrium limit of this novel prescription), without the need to consider the next to leading order.

%%%%%%%%%%

The paper is organized as follows: In the next Section we introduce the hydrodynamics expansion and discuss the anisotropic case, including the generalized anisotropic matching procedure. The selection of the dynamic equations  is discussed in Sec.~\ref{sect:dynamics}. Section~\ref{sect:RTA} contains the analysis of the specific case of the collisional kernel treated in the relaxation-time approximation, the close-to-equilibrium limit, and the numerical results. Summary and conclusions are presented in the last Section. Throughout the paper the natural units are in use, where $c=\hbar=k_B=1$, and the metric tensor has the signature $(+,-,-,-)$. The contraction of vectors (and rank two tensors) can be expressed both by repeated indices in the upper and lower position $U_\mu T^{\mu\nu} U_\nu$ and by a dot $U\cdot T\cdot U$.  Details of the calculations for the $0+1$ dimensional case are presented in the Appendix.

%%%%%%%%%%%%%%%%%%%%%%%%%%%%%%%%%%%%%%%%%%%%%%%%%%%%%%%%%%%%%%%%%%%%%%%%%%%%%%%%%%%%%%%%%%%%%%%%%%%%
\section{Hydrodynamics expansion}
\label{sect:exp}
%%%%%%%%%%%%%%%%%%%%%%%%%%%%%%%%%%%%%%%%%%%%%%%%%%%%%%%%%%%%%%%%%%%%%%%%%%%%%%%%%%%%%%%%%%%%%%%%%%%%%

The purpose of this work is to extract hydrodynamics from the relativistic Boltzmann equation. For mathematical simplicity we consider only a single fluid of massive particles, without external fields, in flat space-time. Therefore, the Boltzmann equation reads

\begin{equation}
 p^\mu\partial_\mu f(x,p) = -\coll[f],
 \label{BE}
\end{equation}
where $f(x,p)$ is the particle distribution and $\coll[f]$ is the collisional kernel, in the most general case a functional of the distribution $f$. Using the kinematic definition of the energy-momentum tensor we can write the local four-momentum conservation as the first moment of the Boltzmann equation~(\ref{BE}). Indeed, with the energy-momentum tensor $T^{\mu\nu}$ defined as
\begin{equation}
 T^{\mu\nu}(x) = \int dP \, p^\mu p^\nu \, f(x,p),
 \label{kineticT}
\end{equation}
the first moment reduces to the ordinary divergence-free condition on the energy-momentum tensor
\begin{equation}
 0= \int dP\,  p^\nu \coll[f] = \int dP \, p^\nu p\cdot\partial f = \partial_\mu \int dP \, p^\mu p^\nu \, f = \partial_\mu T^{\mu\nu}.
 \label{first_moment_0}
\end{equation}
We understand that $\int dP$ in the last formulas is the Lorentz invariant momentum measure $\int dP = \int d^3{\bf p}/E_{\bf p}$, with $E_{\bf p} = \sqrt{m^2 + {\bf p}^2}$ the on-shell energy of particles with mass $m$.

Equation~(\ref{first_moment_0}) is the fundamental equation of hydrodynamics, governing macroscopic evolution of energy and momentum densities\footnote{In presence of additional conserved charges like the baryon number, one should add the relative equations, but in the present paper we consider the simplest case.}. Unfortunately~(\ref{first_moment_0}) contains only four independent equations out of the ten degrees of freedom of the energy-momentum tensor, therefore it cannot be solved directly.

%%%%%%%%%%%%%%%%%%%%%%%%%%%%%%%%%%%%%%%%%%%%%%%%%%%%%%%%%%%%%%%%%%%%%%%%%%%%%%%%%%%%%%%%%%%%%%%%%%%%
\subsection{Viscous hydrodynamics expansion}
\label{sect:aniso}
%%%%%%%%%%%%%%%%%%%%%%%%%%%%%%%%%%%%%%%%%%%%%%%%%%%%%%%%%%%%%%%%%%%%%%%%%%%%%%%%%%%%%%%%%%%%%%%%%%%%%

Traditionally, in order to extract viscous hydrodynamics from an underlying kinetic theory, that is, in order to find the necessary extra equations needed to close the system, one expands the distribution function around the local equilibrium
\begin{equation}
 f(x,p) = f_{\rm eq.}(x,p) + \deltaf(x,p).
 \label{feq}
\end{equation}
For Boltzmann-like statistics, the local equilibrium distribution reads
\begin{equation}
 f_{\rm eq.}(x,p) = k \, \exp\left[ -\frac{p_\mu U^\mu(x)}{T(x)} \right],
 \label{feq}
\end{equation}
which is the point dependent form of the global equilibrium distribution. It is assumed at this stage that the deviation from equilibrium is small and it can be treated in a perturbative manner. For a detailed discussion about the meaning of a perturbative treatment of $\deltaf$, and about different approaches to obtain these extra equations see Ref.~\cite{Denicol:2014loa}.

In the formula~(\ref{feq}) we have not yet defined the fluid four-velocity $U^\mu$ and the effective temperature $T$. A common definition for $U^\mu$ is the Landau definition\footnote{Another popular definition is the Eckart one. It must be noted that these are not the only two possible definitions (see, for instance, Refs.~\cite{Van:2013sma,Becattini:2014yxa,Van:2011yn} for extended discussion).}. The four-velocity is the the time-like eigenvector of the energy-momentum tensor 
\begin{equation}
 U_\mu T^{\mu\nu} = \ped \,  U^\nu,
 \label{Landau_frame_0}
\end{equation}
where $\ped$ is the proper energy density, namely the energy density seen from the local rest frame. In other words, the requirement in Eq.~(\ref{Landau_frame_0}) is equivalent to the condition on $\deltaf$
\begin{equation}
 \Delta^\mu_\alpha U_\beta \, T^{\alpha\beta} = \int dP \, (p\cdot U) p^{\langle \mu \rangle} \, \deltaf =0.
 \label{Landau_frame_1}
\end{equation}
With the notation ${\sf A}^{\dots\langle\mu\rangle\dots} = \Delta^\mu_\nu \,  {\sf A}^{\dots \nu\dots}$ we are indicating the projection orthogonal  to the four-velocity for the index $\mu$, with $\Delta^{\mu\nu}$ being the projector
\begin{equation}
 \Delta^{\mu\nu} = g^{\mu\nu} - U^\mu U^\nu.
 \label{Delta}
\end{equation}
The usual definition of the effective temperature is that it gives the correct proper energy density
\begin{equation}
 \ped = U_\mu U_\nu \, T^{\mu\nu} = \int dP \, (p\cdot U)^2 \, f = \int dP (p\cdot U)^2 f_{\rm eq.},
 \label{LM_0}
\end{equation}
which is generally referred to as the Landau matching. This is implicitly an additional constraint on the possible forms  of $\deltaf$, before making any approximations, namely
\begin{equation}
 \int dP \, (p\cdot U)^2 \, \deltaf = 0.
 \label{LM_1}
\end{equation}
We can present at this point the most general decomposition of the stress-energy tensor consistent with the Landau prescription for the four-velocity
\begin{equation}
 T^{\mu\nu} = \ped \, U^\mu U^\nu - \left( \frac{}{} \preseq + \Pi \right)\Delta^{\mu\nu} + \pi^{\mu\nu}.
 \label{general_decomposition}
\end{equation}
In this notation $\pi^{\mu\nu}$ is the space-like traceless part of $T^{\mu\nu}$, while the isotropic pressure is the "traceful" part. The latter is usually divided into the hydrostatic pressure $\preseq$, that is the pressure that the system would have at global equilibrium having $\ped$ as the energy density, and the bulk viscous correction $\Pi$. Therefore
\begin{equation}
 \pi^{\mu\nu} = T^{\langle\mu\nu\rangle} , \qquad \qquad \Pi = -\frac{1}{3}\Delta_{\mu\nu}T^{\mu\nu} - \preseq,
\end{equation}
with the angular bracket we understand the contraction with $\Delta^{\mu\nu}_{\alpha\beta}$, defined  as
\begin{equation}
 \Delta^{\mu\nu}_{\alpha\beta} = \frac{1}{2} \left( \Delta^\mu_\alpha \Delta^\nu_\beta + \Delta^\nu_\alpha \Delta^\mu_\beta -  \frac{2}{3}\Delta^{\mu\nu} \Delta_{\alpha\beta} \right),
\end{equation}
that is, projecting two indices on the spatial, symmetric and traceless part.

It must be noted that both $\ped$ and $\preseq$ are given solely by the local equilibrium background of the distribution, while the pressure corrections $\pi^{\mu\nu}$ and $\Pi$ depend only on $\deltaf$,
\begin{equation}
 \Pi= -\frac{1}{3}\int dP \, (p\cdot\Delta\cdot p)\deltaf, \qquad \qquad \pi^{\mu\nu} = \int dP \, p^{\langle\mu}p^{\nu\rangle} \, \deltaf,
\end{equation}
therefore, the local equilibrium distribution does not contain any information on the pressure corrections, even if it reproduces  the full proper energy density.

More generally, the decomposition~(\ref{general_decomposition}) is an organization of the ten independent degrees of freedom of the symmetric tensor $T^{\mu\nu}$, namely the proper energy density $\ped$ (and hence the hydrostatic pressure $\preseq$, depending only on $\ped$), three independent components of the time-like vector $U^\mu$, five independent components of $\pi^{\mu\nu}$ and the bulk pressure $\Pi$. Since this decomposition is completely general, it can be used for the anisotropic background too. 

%%%%%%%%%%%%%%%%%%%%%%%%%%%%%%%%%%%%%%%%%%%%%%%%%%%%%%%%%%%%%%%%%%%%%%%%%%%%%%%%%%%%%%%%%%%%%%%%%%%%
\subsection{Anisotropic expansion}
\label{sect:aniso}
%%%%%%%%%%%%%%%%%%%%%%%%%%%%%%%%%%%%%%%%%%%%%%%%%%%%%%%%%%%%%%%%%%%%%%%%%%%%%%%%%%%%%%%%%%%%%%%%%%%%%

The background of the hydrodynamics expansion does not need to be the local equilibrium. In anisotropic hydrodynamics one expands the distribution function around an anisotropic background embedding (at least partially) the pressure corrections characterizing the full system
\begin{equation}
 f(x,p) = f_a(x,p) + \tdeltaf.
\end{equation}
In the previous works~\cite{Bazow:2013ifa,Bazow:2015cha, Tinti:2014yya} it was assumed that the pressure corrections are not fully described by the anisotropic background $f_a$, and in general there should be some (expected to be small) contributions from $\tdeltaf$. However, in the present work we propose a different approach, extending the Landau matching to the anisotropic case.

Because of the success of previous applications of anisotropic hydrodynamics, let us consider, for instance, the distribution
\begin{equation}
 f_a(x,p) = k \exp\left[ -\frac{1}{\Lambda(x)} \sqrt{\frac{}{}p_\mu \Xi^{\mu\nu}(x) p_\nu}\right],
 \label{fa_0}
\end{equation}
which is a generalization of the Romatschke-Strickland form~\cite{Romatschke:2003ms}. It must be noted that the form of the anisotropic background~(\ref{fa_0}) is not the only one possible. The extended matching procedure we advocate in this Section can be applied to different backgrounds. However, in the present work we consider only this background, assuming that the generalized Romatschke-Strickland form provides a good enough approximation of the full distribution function and deviations from (\ref{fa_0}) are small perturbations even if there are large gradients in the system.

The most general decomposition of the symmetric tensor $\Xi^{\mu\nu}$, compatible with the Landau prescription of the four-velocity, is
\begin{equation}
 \Xi^{\mu\nu} = \phi_U\, U^\mu U^\nu + \phi_s \Delta^{\mu\nu} + \Xi^{\langle\mu\nu\rangle}.
\end{equation}
It is convenient now to use the definition of the projector $\Delta^{\mu\nu}$ in~(\ref{Delta}) to write
\begin{equation}
 \Xi^{\mu\nu} = \left( \frac{}{} \phi_U-\phi_s \right)U^\mu U^\nu + \phi_s g^{\mu\nu} + \Xi^{\langle\mu\nu\rangle}.
\end{equation}
In order to always have a real valued $f_a$, taking into account that particles are on-shell, both $\phi_U$ and $\phi_U-\phi_s$ must be positive. Therefore, in a completely general way, it is possible to rescale the quantities and write
\begin{equation}
 f_a (x,p) = k \exp \left[ -\frac{1}{\lambda}\sqrt{ p_\mu \left( \frac{}{} U^\mu U^\nu + \phi \, g^{\mu\nu} +\xi^{\mu\nu} \right)p_\nu } \right] = k  \exp \left[ -\frac{1}{\lambda}\sqrt{ m^2 \phi  + p_\mu \left( \frac{}{} U^\mu U^\nu+\xi^{\mu\nu} \right)p_\nu } \right] ,
 \label{fa_1}
\end{equation}
where
\begin{equation}
 \frac{1}{\lambda^2} = \frac{\phi_u-\phi_s}{\Lambda^2}, \qquad \phi = \frac{\phi_s}{\phi_U-\phi_s}, \qquad \xi^{\mu\nu} = \frac{\Xi^{\langle \mu\nu \rangle}}{\phi_U-\phi_s}.
\end{equation}
In Refs~\cite{Nopoush:2014nc,Tinti:2014yya} it was preferred to have the $\phi$ scalar multiplying the projector $\Delta^{\mu\nu}$. This can be easily done here by rescaling the quantities in~(\ref{fa_0}) around $\phi_U$ instead of $\phi_U -\phi_s$. In this paper the prescription~(\ref{fa_1}) is adopted for mathematical convenience, the reason will become apparent later. We note that we have now a very similar situation to the previous case. Indeed, we recover the usual local equilibrium distribution~(\ref{feq}) in the limit of vanishing anisotropy tensor $\xi^{\mu\nu}$, vanishing $\phi$ and, therefore, $\lambda$ becomes the effective temperature $T$.

Having fixed the form of the background distribution, we should clearly specify the independent degrees of freedom. For the local equilibrium distribution~(\ref{feq}), it was necessary to define four parameters, the effective temperature $T$ and three independent components of the four velocity field $U^\mu$. Now we aim to generalize the procedure to the anisotropic background~(\ref{fa_1}). The four-velocity can be defined adopting the Landau prescription~(\ref{Landau_frame_0}), exactly as in the previous case. Therefore $\tdeltaf$ must fulfill a condition which is equivalent to the one seen for the viscous expansion~(\ref{Landau_frame_1}). Unfortunately there is no direct way to fix each parameter separately. However, looking at the Landau matching~(\ref{LM_0}) for the definition of the temperature, the most straightforward extension is to fix the remaining parameters in such a way that the leading order integrals match with the exact values of the whole stress-energy tensor. 

The Landau prescription takes care of  the three degrees of freedom for the four--velocity, hence,  there are seven remaining independent degrees of freedom, namely the momentum scale $\lambda$, the bulk parameter $\phi$ \footnote{It will be shown later that the bulk viscosity $\Pi$ is proportional to $\phi$ at first order in small deviations from local equilibrium.} and five independent components of the anisotropy tensor $\xi^{\mu\nu}$. This is the same number of effective degrees of freedom as in the general decomposition~(\ref{general_decomposition}), where we have: the four velocity $U^\mu$,  the proper energy density $\ped$, the isotropic pressure $\preseq + \Pi$ (and hence $\Pi$, since $\preseq$ is already defined by $\ped$ if there are no additional conserved charges), and $\pi^{\mu\nu}$. Therefore, as a generalization of the Landau matching to the anisotropic case, we are assuming that all of these quantities will be reproduced by the anisotropic background alone
\begin{equation}
 T^{\mu\nu} = \int dP \, p^\mu p^\nu \, f = \int dP \, p^\mu p^\nu \, f_a
 \label{aLM_0}
\end{equation}
or, equivalently
\begin{equation}
 \int dP \, p^\mu p^\nu \, \tdeltaf = 0,
 \label{aLM_1}
\end{equation}
which is a generalization of Eqs.~(\ref{Landau_frame_1}) and~(\ref{LM_1}) to the anisotropic background. The reason of the choice of  the parametrization~(\ref{fa_1}) is clear now. In the massless case the system is exactly conformal and the energy-momentum tensor is traceless, $T^\mu_\mu=0$. Thus, there is one less degree of freedom in $T^{\mu\nu}$. On the other hand, in the anisotropic distribution~(\ref{fa_1}), the bulk degree of freedom $\phi$ vanishes too because of the $m^2$ factor multiplying it. Hence, we still have the same number of degrees of freedom in the anisotropic background as in the exact energy-momentum tensor.

%%%%%%%%%%%%%%%%%%%%%%%%%%%%%%%%%%%%%%%%%%%%%%%%%%%%%%%%%%%%%%%%%%%%%%%%%%%%%%%%%%%%%%%%%%%%%%%%%%%%
\section{Dynamical equations}
\label{sect:dynamics}
%%%%%%%%%%%%%%%%%%%%%%%%%%%%%%%%%%%%%%%%%%%%%%%%%%%%%%%%%%%%%%%%%%%%%%%%%%%%%%%%%%%%%%%%%%%%%%%%%%%%%

In this Section we will introduce the equations of motions for the anisotropic expansion. We follow mainly the approach of Refs.~\cite{Denicol:2012cn,Jaiswal:2013npa}, generalizing the arguments to the anisotropic expansion. The main idea is to get the equations directly from the exact kinematic evolution, then making an approximation on $\tdeltaf$ (or its moments) in order to have a finite number of equations.

In order to have energy and momentum conservation, it is necessary to take the first moment of the Boltzmann equation~(\ref{first_moment_0}). Making use of~(\ref{general_decomposition}), and taking the projection parallel and orthogonal to the four-velocity, one can write

\begin{eqnarray}\label{e_ev}
 && D\ped = -\left( \frac{}{} \ped +\preseq + \Pi \right)\theta + \sigma_{\mu\nu}\pi^{\mu\nu}, \\ \label{U_ev}
 && \left( \frac{}{}\ped +\preseq +\Pi \right)D U^\alpha = -\nabla^\alpha\left( \frac{}{}\preseq + \Pi \right) -\Delta^\alpha_\mu\partial_\nu \pi^{\mu\nu},
\end{eqnarray}
where we used the the notation for the convective derivative $D = U^\mu \partial_\mu$,  and for the spatial gradient $\nabla^\mu = \Delta^{\mu\nu}\partial_\nu$. The scalar expansion $\theta = \partial_\mu U^\mu$ and the shear stress $\sigma^{\mu\nu} = \Delta^{\mu\nu}_{\alpha\beta} \, \partial^\alpha U^\beta$ are the ones appearing in the familiar decomposition of the expansion tensor $\theta^{\mu\nu}$
\begin{equation}
 \theta^{\mu\nu} = \nabla^\mu U^\nu = \sigma^{\mu\nu} + \theta\Delta^{\mu\nu} + \omega^{\mu\nu},
\end{equation}
with $\omega^{\mu\nu} = \nabla^\mu U^\nu -\nabla^\nu U^\mu $ being the vorticity.

Containing only one convective derivative, Eqs.~(\ref{e_ev}) and~(\ref{U_ev}) are usually considered the time evolution equations for the proper energy density $\ped$ and the four-velocity $U^\mu$. It is possible to obtain from kinetic theory a set of equations for the remaining degrees of freedom, however, in general it is not guaranteed that the convective derivative will appear only acting on $\pi^{\mu\nu}$ and $\Pi$. Using the orthogonality properties of the projectors $\Delta^{\mu\nu}_{\alpha\beta}$, $\Delta^\mu_\nu$ and the Landau prescription for the four-velocity~(\ref{Landau_frame_0}), the {\it exact} convective derivatives of the pressure corrections read
\begin{equation}
 D\pi^{\langle\mu\nu\rangle} = \int dP \, p^{\langle\mu}p^{\nu\rangle} \, Df, \qquad D\Pi = -\frac{1}{3}\int dP \left( \frac{}{}p\cdot\Delta\cdot p \right) Df - D\preseq.
 \label{ev_pres_corr}
\end{equation}
It is interesting to note at this point that the four-momentum conservation equations and the last equations put together are just another way to rewrite the {\it exact} evolution equation of all the components of the energy-momentum tensor
\begin{equation}\label{Tmunu_ev}
 D T^{\mu\nu} = \int dP \, p^\mu p^\nu \, D f.
\end{equation}
Indeed, making use of the general decomposition~(\ref{general_decomposition}) and the local four-momentum conservation constraint of the collisional kernel $\int dP \, p^\mu {\cal C}[f] =0$, it is straightforward to prove that the $U_\mu U_\nu$ contraction correspond to Eq.~(\ref{e_ev}), the $\Delta^\alpha_\mu U_\nu$ one corresponds to Eq.~(\ref{U_ev}), and the remaining independent ones, namely the spatial traceless part and the spatial trace of Eq.~(\ref{Tmunu_ev}), correspond respectively to Eqs.~(\ref{ev_pres_corr})\footnote{Note that no expansion of the distribution function has been used so far, the arguments are completely general.}.

It is convenient now to write the evolution of pressure corrections in terms of moments of the distribution function instead of moments of the derivatives of the distribution function. From the Boltzmann equation~(\ref{BE}) it is possible to obtain the {\it exact} convective derivative of the distribution function
\begin{equation}
 D f = \frac{1}{(p\cdot U)}\left[ (p\cdot U)U^\mu \partial_\mu f = p\cdot\partial f - p\cdot \nabla f \frac{}{}\right] = -\frac{\coll[f]}{(p\cdot U)} -\frac{p\cdot\nabla f}{(p\cdot U)}.
\end{equation}
Making use of the notation
\begin{equation}
 \coll^{\mu_1 \dots \mu_n}_r = \int dP \, (p\cdot U)^r \, p^{\mu_1}\dots p^{\mu_n} \coll[f],
\end{equation}
the dynamical equations~(\ref{ev_pres_corr}) read
\begin{eqnarray}\label{shear}
  D\pi^{\langle\mu\nu\rangle} +\coll^{\langle\mu\nu\rangle}_{-1} = -\Delta^{\mu\nu}_{\rho\sigma} \nabla_\alpha  \int dP \,\frac{ p^\rho p^\sigma p^\alpha  \,  f}{(p\cdot U)} \, -  \left( \frac{}{}\sigma_{\rho \sigma} +\frac{1}{3} \, \theta \, \Delta_{\rho\sigma} \right) \int dP \,\frac{ p^{\langle\mu} p^{\nu\rangle} p^\rho p^\sigma  \,  f}{(p\cdot U)^2}, \\\label{bulk}
D \Pi - \frac{1}{3} \Delta_{\mu\nu}\coll^{\mu\nu}_{-1} =-D\preseq + \frac{1}{3}\Delta_{\mu\nu} \nabla_\rho  \int dP \,\frac{ p^\mu p^\nu p^\rho  \,  f}{(p\cdot U)} \, + \frac{1}{3} \left( \frac{}{}\sigma_{\rho \sigma} +\frac{1}{3} \, \theta \, \Delta_{\rho\sigma} \right)  \int dP \,\frac{ (p\cdot \Delta \cdot p) p^\rho p^\sigma  \,  f}{(p\cdot U)^2}.
\end{eqnarray}
It is important to note at this point that Eqs.~(\ref{shear}) and~(\ref{bulk}) do not depend only on the background $f_a$. In fact, even if the second moment of the full distribution $f$ does not depend on the correction $\tdeltaf$ because of~(\ref{aLM_1}), the same cannot be said of the other moments of the distribution (including possible powers of $(p\cdot U)$). Indeed, both the integrals of the collisional term on the left-hand side of (\ref{shear}) and (\ref{bulk}), and the integrals on the right-hand side cannot be reduced to a function of the elements of $T^{\mu\nu}$ in the most general case. Equations~(\ref{e_ev}),~(\ref{U_ev}), ~(\ref{shear}) and~(\ref{bulk}) can form a close system of equations only at the leading order of the anisotropic expansion, that is, neglecting the contribution of $\tdeltaf$.

In order to have a more accurate description, it is possible to use the same procedure we used for obtaining Eq.~(\ref{Tmunu_ev}) (namely, four-momentum conservation plus  Eqs.~(\ref{shear}) and~(\ref{bulk})) to all of the other moments of the distribution function 

\begin{eqnarray} \nonumber
D \int dP \, (p\cdot U)^r \, p^{\mu_1}\dots p^{\mu_n} \, f &=& r \, DU_\nu \int dP (p\cdot U)^{r-1}  \, p^{\mu_1}\dots p^{\mu_n} p^\nu \, f  -\coll_{r-1}^{\mu_1\dots\mu_n} - \nabla_\nu \int dP\, (p\cdot U)^{r-1} \, p^{\mu_1}\dots p^{\mu_n} \, f \\ \label{exact_moments}
&&  + \left(  r-1 \right) \left( \frac{}{}\sigma_{\rho \sigma} +\frac{1}{3} \, \theta \, \Delta_{\rho\sigma}  \right)\int dP \, (p\cdot U)^{r-2} \, p^{\mu_1}\dots p^{\mu_n} p^\rho p^\sigma \, f .
\end{eqnarray}
This unfortunately provides an infinite set of coupled equations, since the equation for any moment  contains higher moments of the distribution function. This is not unexpected, indeed, there is exactly the same problem in the ordinary viscous expansion~\cite{Denicol:2012cn} around the local equilibrium, once we take into account the different notation. The usual procedure is to express the deviation from the background $\tdeltaf$ explicitly as series of irreducible moments of $\tdeltaf$ itself, truncating the series in order to have a finite number of degrees of freedom. See for instance Ref.~\cite{Denicol:2014loa}, for an extended discussion of different truncation schemes, including different choices of dynamical equations instead of the exact moments~(\ref{exact_moments}) for closing the set of equations. The expansion in Ref.~\cite{Denicol:2014loa} is around local equilibrium, but the arguments can be straightforwardly extended to the anisotropic background. In the specific case of anisotropic hydrodynamics, Bazow at al.~\cite{Bazow:2013ifa,Bazow:2015cha} have been using an ``Israel-Stewart like'' expansion for $\tdeltaf$. For the reminder of this work we will not address the most convenient truncation for the deviation from the background, leaving it for further research. Starting from the next Section we will consider only the leading order of the anisotropic expansion, namely $f_a$.

As a final note, as long as we consider the leading order of anisotropic hydrodynamics, the condition~(\ref{aLM_1}) is not essential, since $\tdeltaf$ is systematically neglected. We could still use Eqs.~(\ref{e_ev}),~(\ref{U_ev}),~(\ref{shear}) and~(\ref{bulk}) for closing the system of equations. However, in order for the next to leading order to stem from the exact evolution of the moments of the full distribution~(\ref{exact_moments}) like the leading order, it is mandatory. In fact in Refs.~\cite{Bazow:2013ifa,Bazow:2015cha,Tinti:2014yya}, only the Landau prescription~(\ref{Landau_frame_1}) and the Landau matching~(\ref{LM_1}) were considered, instead of the full generalized matching~(\ref{aLM_1}). The unfortunate consequence is that Eqs.~(\ref{shear}) and~(\ref{bulk}) can not be used to close the system of equations both for the degrees of freedom of the anisotropic distribution function and the extra corrections from $\tdeltaf$ . Lacking any clear prescription to find the extra needed equations, it comes back the problem of ambiguity in the derivation of fluid dynamics already seen in viscous hydrodynamics (see the discussion on Israel-Stewart approach in Ref.~\cite{Denicol:2014loa}). For instance, in Ref.~\cite{Bazow:2013ifa,Bazow:2015cha} the needed extra equations for closing viscous anisotropic hydrodynamics have been selected from the zeroth and second moment of the Boltzmann equation.

%%%%%%%%%%%%%%%%%%%%%%%%%%%%%%%%%%%%%%%%%%%%%%%%%%%%%%%%%%%%%%%%%%%%%%%%%%%%%%%%%%%%%%%%%%%%%%%%%%%%
\section{Specific example: leading order for the relaxation-time approximation}
\label{sect:RTA}
%%%%%%%%%%%%%%%%%%%%%%%%%%%%%%%%%%%%%%%%%%%%%%%%%%%%%%%%%%%%%%%%%%%%%%%%%%%%%%%%%%%%%%%%%%%%%%%%%%%%%

In this Section we discuss the specific case of the collisional kernel $\coll[f]$ treated in the relaxation-time approximation (RTA)
\begin{equation}
 \coll[f] = \frac{1}{\taueq}(p\cdot U)\left[ \frac{}{} f(x,p) - \feq(x,p) \right],
 \label{RTA}
\end{equation}
with  $\taueq$ being the relaxation time and $\feq$ the local equilibrium distribution~(\ref{feq})\footnote{The local equilibrium distribution is well defined in any situation through the four velocity prescription~(\ref{Landau_frame_0}) and the Landau matching~(\ref{LM_0}), even if it is not used as the background in the hydrodynamics expansion. }.

Thanks to Eq.~(\ref{RTA}) the collisional kernel in Eqs.~(\ref{shear}) and~(\ref{bulk}) can be easily calculated, and we can write
\begin{eqnarray}\label{shear_RTA_0}
  D\pi^{\langle\mu\nu\rangle} + \frac{1}{\taueq} \pi^{\mu\nu} = -\Delta^{\mu\nu}_{\rho\sigma} \nabla_\alpha  \int dP \,\frac{ p^\rho p^\sigma p^\alpha  \,  f}{(p\cdot U)} \, -  \left( \frac{}{}\sigma_{\rho \sigma} +\frac{1}{3} \, \theta \, \Delta_{\rho\sigma}  \right) \int dP \,\frac{ p^{\langle\mu} p^{\nu\rangle} p^\rho p^\sigma  \,  f}{(p\cdot U)^2}, \\\label{bulk_RTA_0}
D \left( \frac{}{} \Pi +\preseq\right) + \frac{1}{\taueq}\Pi = + \frac{1}{3}\Delta_{\mu\nu} \nabla_\rho  \int dP \,\frac{ p^\mu p^\nu p^\rho  \,  f}{(p\cdot U)} \, + \frac{1}{3} \left( \frac{}{}\sigma_{\rho \sigma} +\frac{1}{3} \, \theta \, \Delta_{\rho\sigma} \right)  \int dP \,\frac{ (p\cdot \Delta \cdot p) p^\rho p^\sigma  \,  f}{(p\cdot U)^2}.
\end{eqnarray}
We are interested in the leading order of the expansion, therefore we will consider the approximation $f\simeq f_a$ with the anisotropic distribution function defined in~(\ref{fa_1}). Because of the special form of the collisional kernel in~(\ref{RTA}) and the extended matching~(\ref{aLM_1}), the left-hand sides of both Eq.~(\ref{shear_RTA_0}) and Eq.~(\ref{bulk_RTA_0}) do not depend on $\tdeltaf$. On the other hand, the first term on the right-hand sides of both equations can be written directly in terms of the components of $T^{\mu\nu}$. Indeed, because of the symmetry of $f_a$\footnote{In fact if $\tdeltaf$ preserves the parity invariance (seen from the local rest frame) of $f_a$, this is still valid. However in the most general case there are some other non vanishing terms. }
\begin{equation}
 \int dP \,\frac{ p^\rho p^\sigma p^\alpha  \,  f_a}{(p\cdot U)} = U^\rho T^{\sigma\alpha} + U^\sigma T^{\alpha\rho} + U^\alpha T^{\rho\sigma} -2 \, \ped \, U^\rho U^\sigma U^\alpha,
\end{equation}
therefore, Eqs.~(\ref{shear_RTA_0}) and~(\ref{bulk_RTA_0}) read
\begin{eqnarray}\label{shear_RTA_LO_0}
  D\pi^{\langle\mu\nu\rangle} + \frac{1}{\taueq} \pi^{\mu\nu} &=&  -  \left( \frac{}{}\sigma_{\rho \sigma} +\frac{1}{3} \, \theta \, \Delta_{\rho\sigma}  \right) \int dP \,\frac{ p^{\langle\mu} p^{\nu\rangle} p^\rho p^\sigma  \,  f_a}{(p\cdot U)^2} - 2 \, \pi_\alpha^{<\mu}\sigma^{\nu>\alpha} \\ \nonumber
  && \quad + 2\left( \frac{}{} \preseq + \Pi \right)\sigma^{\mu\nu} -\frac{5}{3}\, \theta \, \pi^{\mu\nu} + 2 \, \pi_\alpha^{<\mu}\omega^{\nu>\alpha}, \\ \nonumber \\ \label{bulk_RTA_LO_0}
D \left( \frac{}{} \Pi +\preseq \right) + \frac{1}{\taueq}\Pi &=&  \frac{1}{3} \left( \frac{}{}\sigma_{\rho \sigma} +\frac{1}{3} \, \theta \, \Delta_{\rho\sigma} \right)  \int dP \,\frac{ (p\cdot \Delta \cdot p) p^\rho p^\sigma  \,  f_a}{(p\cdot U)^2} \\ \nonumber
&& \quad +\frac{2}{3} \, \pi_{\mu\nu}\sigma^{\mu\nu} - \frac{5}{3} \left( \preseq +\Pi \right) \theta.
\end{eqnarray}
It is convenient at this point to introduce the notation for the isotropic pressure
\begin{equation}
 \pres = \preseq + \Pi = -\frac{1}{3}\int dP \, \left( p\cdot\Delta\cdot p \right) f_a,
\end{equation}
since, differently from the hydrostatic pressure $\preseq$ itself, it does not depend on the equilibrium distribution function $\feq$. Equations~(\ref{shear_RTA_LO_0}) and~(\ref{bulk_RTA_LO_0}) then read
\begin{eqnarray}\label{shear_RTA_LO_1}
  D\pi^{\langle\mu\nu\rangle} + \frac{1}{\taueq} \pi^{\mu\nu} &=&  -  \left( \frac{}{}\sigma_{\rho \sigma} +\frac{1}{3} \, \theta \, \Delta_{\rho\sigma}  \right) \int dP \,\frac{ p^{\langle\mu} p^{\nu\rangle} p^\rho p^\sigma  \,  f_a}{(p\cdot U)^2} - 2 \, \pi_\alpha^{<\mu}\sigma^{\nu>\alpha} \\ \nonumber
  && \quad + 2\, \pres \, \sigma^{\mu\nu} -\frac{5}{3}\, \theta \, \pi^{\mu\nu} + 2 \, \pi_\alpha^{<\mu}\omega^{\nu>\alpha}, \\ \nonumber \\ \label{bulk_RTA_LO_1}
D\pres + \frac{1}{\taueq}\left( \frac{}{} \pres-\preseq \right) &=&  \frac{1}{3} \left( \frac{}{}\sigma_{\rho \sigma} +\frac{1}{3} \, \theta \, \Delta_{\rho\sigma} \right)  \int dP \,\frac{ (p\cdot \Delta \cdot p) p^\rho p^\sigma  \,  f_a}{(p\cdot U)^2} \\ \nonumber
&& \quad +\frac{2}{3} \, \pi_{\mu\nu}\sigma^{\mu\nu} - \frac{5}{3} \, \pres \,  \theta.
\end{eqnarray}
%
%Here, the only term depending on the local equilibrium distribution function is no longer appearing on derivatives.

%%%%%%%%%%%%%%%%%%%%%%%%%%%%%%%%%%%%%%%%%%%%%%%%%%%%%%%%%%%%%%%%%%%%%%%%%%%%%%%%%%%%%%%%%%%%%%%%%%%%
\subsection{Close-to-equilibrium limit}
\label{sect:linearized}
%%%%%%%%%%%%%%%%%%%%%%%%%%%%%%%%%%%%%%%%%%%%%%%%%%%%%%%%%%%%%%%%%%%%%%%%%%%%%%%%%%%%%%%%%%%%%%%%%%%%%

It is well known that second order viscous hydrodynamics is a very good approximation if the system is close to local equilibrium. Therefore, if the prescription we present for anisotropic hydrodynamics is a good description of the system, we expect that our equations should match with the second order viscous hydrodynamics in the close-to-equilibrium limit. Namely, Eqs.~(\ref{shear_RTA_LO_0}) and~(\ref{bulk_RTA_LO_0}) should reduce to\footnote{Note that in general there can be more second order terms, we are only showing the non vanishing ones in the close-to-equilibrium limit of anisotropic hydrodynamics.}
\begin{eqnarray}
 \tau_\pi D\pi^{\langle\mu\nu\rangle} + \pi^{\mu\nu} &=& 2\eta \, \sigma^{\mu\nu} + 2 \, \tau_\pi \, \pi_\lambda^{\langle\mu}\omega^{\nu\rangle\lambda} -\tau_{\pi\pi}  \: \pi_\lambda^{\langle\mu}\sigma^{\nu\rangle\lambda} - \delta_{\pi\pi} \, \pi^{\mu\nu}\theta + \lambda_{\pi\Pi} \, \Pi \, \sigma^{\mu\nu} ,
\label{visc_shear} \\
\tau_\Pi \, D\Pi +  \Pi &=&  -\zeta \, \theta  - \delta_{\Pi\Pi} \, \Pi \, \theta + \lambda_{\Pi\pi} \, \pi^{\mu\nu}\sigma_{\mu\nu} ,
\label{visc_bulk}
\end{eqnarray}
with certain values of the transport coefficients.

In the previous works on anisotropic hydrodynamics it has been proven that the correct close-to-equilibrium behaviour is reproduced, see for instance Refs.~\cite{Martinez:2010sc,Tinti:2014conf,Tinti:2014yya}. To demonstrate this property, it was necessary to make an expansion of the anisotropic background around the local equilibrium, keeping only the lowest order terms, and checking that the equations obtained in this approximation reproduce the second order viscous hydrodynamics. In the present work such checking of the close-to-equilibrium limit is easier, since the equation of motion for the pressure corrections~(\ref{shear_RTA_0}) and~(\ref{bulk_RTA_0}) are exactly the same as the two prescriptions for second order viscous hydrodynamics tested in Ref.~\cite{Jaiswal:2014}. As we show below, the linear expansion of $f_a$ around equilibrium yields exactly the distribution function known from the gradient expansion. Therefore, the transport coefficients in~(\ref{visc_shear}) and~(\ref{visc_bulk}) are exactly the same as those obtained in the Chapman-Enskog-like expansion presented in Ref.~\cite{Jaiswal:2014}, which is in our opinion the best second order viscous hydrodynamics prescription for dealing with the RTA kinetic theory.

Indeed, if we take the expansion of~(\ref{fa_1}) up to the leading order in the deviation from local equilibrium we have
\begin{equation}\label{f->eq}
 f_a \simeq \feq\left[ 1 + \frac{\lambda -T}{T^2}(p\cdot U) - \frac{m^2 \phi}{2T(p\cdot U)} -\frac{1}{2T(p\cdot U)} \, p\cdot \xi \cdot p \right].
\end{equation}
Introducing the notation
\begin{equation}
 I_{n,q} = \frac{1}{(2q+1)!!}\int dP \, (p\cdot U)^{n-2q} (-p\cdot\Delta\cdot p)^q\feq,
\end{equation}
the energy-momentum tensor then reads
\begin{eqnarray}\label{Tmunu_lin_0}
 T^{\mu\nu} &\simeq&  T_{\rm eq.}^{\mu\nu} + \left[ \frac{\lambda-T}{T^2} \, I_{3,0} -\frac{m^2}{2T} \, \phi \, I_{1,0} \right] U^\mu U^\nu - \left[ \frac{\lambda-T}{T^2} \, I_{3,1}-\frac{m^2}{2T} \, \phi  \, I_{1,1} \right] \Delta^{\mu\nu} -\frac{1}{T} \, I_{3,2} \, \xi^{\mu\nu},
\end{eqnarray}
where $T_{\rm eq.}^{\mu\nu}$ stands for
\begin{equation}
 T_{\rm eq.}^{\mu\nu} = \int dP \, p^\mu p^\nu \feq = \ped U^\mu U^\nu -\preseq \Delta^{\mu\nu}.
\end{equation}
Since the energy density cannot have any corrections, because of the definitions of both $\feq$ and $f_a$, we can remove $\lambda$ from Eq.~(\ref{Tmunu_lin_0}). Indeed, since
\begin{equation}\label{LM_lin}
 \frac{\lambda-T}{T^2} \, I_{3,0} = \frac{m^2}{2T} \, \phi \, I_{1,0} \qquad \Rightarrow \qquad \frac{\lambda-T}{T^2}  = \frac{m^2}{2T} \, \phi \, \frac{ I_{1,0}}{I_{3,0}},
\end{equation}
the energy-momentum tensor reads
\begin{eqnarray}\label{Tmunu_lin_1}
 T^{\mu\nu} &\simeq&  T_{\rm eq.}^{\mu\nu} - \frac{m^2}{2T} \, \phi\left[ \frac{ I_{1,0}}{I_{3,0}} \, I_{3,1} -  I_{1,1} \right] \Delta^{\mu\nu} -\frac{1}{T} \, I_{3,2} \, \xi^{\mu\nu}.
\end{eqnarray}
We can recognize then
\begin{align} \label{lin_shear}
 \pi^{\mu\nu} = -\frac{1}{T} \, I_{3,2} \, \xi^{\mu\nu} = -\beta_\pi \, \xi^{\mu\nu}, \\ \label{lin_bulk}
 \Pi = \frac{m^2}{2T} \, \phi \left[ \frac{ I_{1,0}}{I_{3,0}} \, I_{3,1} -  I_{1,1} \right],
\end{align}
where $\beta_\pi$ is the same as the coefficient in Ref.~\cite{Jaiswal:2014}. Making use of the last equations and of Eq.~(\ref{LM_lin}), expansion~(\ref{f->eq}) reads
\begin{equation}\label{fa->CE_0}
 f_a \simeq \feq \left\{ 1 -  \frac{\Pi}{(p\cdot U)\left[ \frac{ I_{1,0}}{I_{3,0}} \, I_{3,1} -  I_{1,1} \right]} \left[ \frac{}{} 1 - \frac{ I_{1,0}}{I_{3,0}} (p\cdot U)^2  \right] + \frac{1}{2T\beta_\pi} \, p\cdot\pi \cdot p \right\}.
\end{equation}
Using the exact relations between the $I_{n,q}$ thermodynamical integrals
\begin{align}
 I_{n,q} = \frac{1}{2q+1}\left[ \frac{}{} I_{n,q-1} -m^2 I_{n-2,q-1} \right], \\
 I_{n,q} = T\left[  I_{n-1,q-1} + \left( \frac{}{} n-2q \right)I_{n-1,q} \right],
\end{align}
the expansion of the anisotropic bakground~(\ref{fa->CE_0}) reads
\begin{equation}\label{fa->CE_1}
 f_a \simeq \feq \left\{ 1 -  \frac{\Pi}{3T(p\cdot U)\beta_\Pi} \left[ \frac{}{} m^2 - \left( 1-3 \, \frac{I_{3,1}}{I_{3,0}} \right)(p\cdot U)^2  \right] + \frac{1}{2T\beta_\pi} \, p\cdot\pi \cdot p \right\},
\end{equation}
where
\begin{equation}
 \beta_\Pi = \frac{5}{3}\beta_\pi - \frac{I_{3,1}}{I_{3,0}} \left( \frac{}{} I_{2,0} + I_{2,1} \right). 
\end{equation}
The right-hand side of~(\ref{fa->CE_1}) matches exactly the form of the distribution function that was plugged in Eqs.~(\ref{shear_RTA_0}) and~(\ref{bulk_RTA_0}) for obtaining the second order viscous hydrodynamics in Ref.~\cite{Jaiswal:2014}, {\it q.e.d.} Naturally, it would have been possible to plug directly the expansion~(\ref{f->eq}) into the right-hand side\footnote{The left hand side is already exact at any order in the expansion of the distribution function. } of Eqs.~(\ref{shear_RTA_LO_1}) and~(\ref{bulk_RTA_LO_1}) and obtain the same results.

%%%%%%%%%%%%%%%%%%%%%%%%%%%%%%%%%%%%%%%%%%%%%%%%%%%%%%%%%%%%%%%%%%%%%%%%%%%%%%%%%%%%%%%%%%%%%%%%%%%%
\subsection{Numerical results}
\label{sect:numerics}
%%%%%%%%%%%%%%%%%%%%%%%%%%%%%%%%%%%%%%%%%%%%%%%%%%%%%%%%%%%%%%%%%%%%%%%%%%%%%%%%%%%%%%%%%%%%%%%%%%%%%

In this Section we present direct comparisons of the new prescription for the anisotropic hydrodynamics expansion with the exact solutions of the Boltzmann equation. We consider the exact solution found in Ref.~\cite{Florkowski:2014sfa}, namely the $0+1$-dimensional solution for a massive gas with the collisional kernel treated in the relaxation-time approximation. We consider a fixed relaxation time $\taueq=0.5$ fm/c, a starting temperature $T_0 = 600$~MeV, and a starting time~$\tau_0 = 0.5$ fm/c. For the mass we consider two options, namely, $m = 0.3$~GeV and $m = 1$~GeV. For the initial anisotropy, we take into account two extreme situations, the system that is initially in local equilibrium, $\xi_0 = 0$, and the system with a very large starting anisotropy, $\xi_0=100$. In Appendix~\ref{sect:explicit} we present the $0+1$-dimensional expressions of the equations presented in this paper. We plot the evolution of the pressure anisotropy, that is the ratio of the longitudinal pressure $\pres_L$ and the transverse pressure $\pres_T$\footnote{See the Appendix for more details.}, and the bulk pressure correction times the proper time $\tau \Pi$.  

%%%%%%%%%%%%%%%%%%%%%%%%%%%%%%%%%%%%%%%%%%%%%%%%%%%%%%%%%%%%%%%%%%%%%%%%%%%%%%%%%%%%%%%%%%
\begin{figure}[t]
\begin{center}
\includegraphics[angle=0,width= \textwidth]{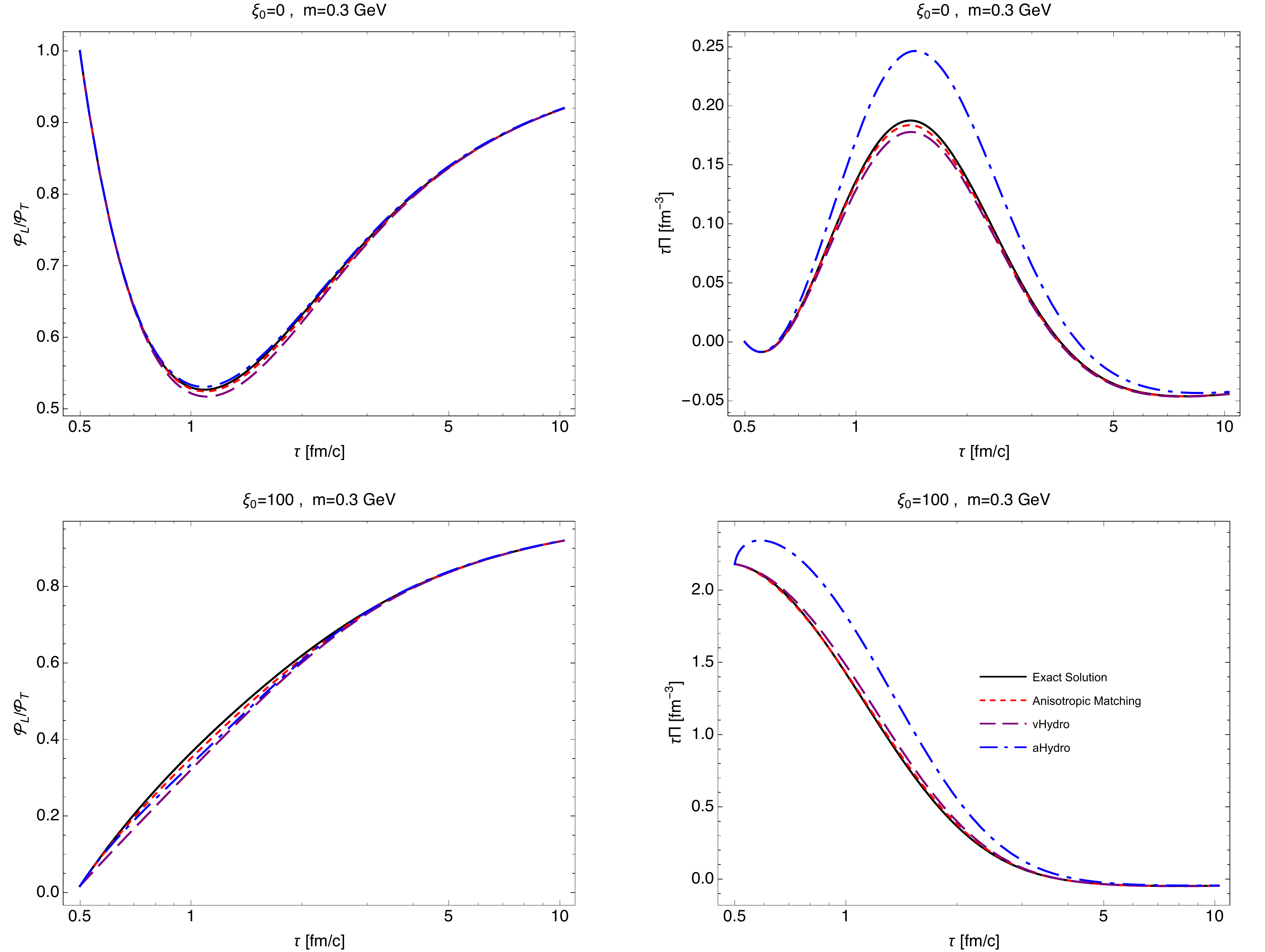}
\end{center}
\caption{(Color online) Time evolution of the pressure anisotropy $\pres_L/\pres_T$ (left) and the bulk corrections time the proper time $\tau \Pi$ (right). The bleck solid line correspond to the exact solution of the Boltzmann equation, the red dotted one to the set of equation presented in this paper, the purple dashed line is the solution of second order viscous hydrodynamics, while the blue dashed-dotted is anisotropic hydrodynamics. The initial anisotropy is vanishing (top) or $\xi_0 = 100$ (bottom) in the two rows.
}
\label{fig:03GeV}
\end{figure}
%%%%%%%%%%%%%%%%%%%%%%%%%%%%%%%%%%%%%%%%%%%%%%%%%%%%%%%%%%%%%%%%%%%%%%%%%%%%%%%%%%%%%%%%%%

%%%%%%%%%%%%%%%%%%%%%%%%%%%%%%%%%%%%%%%%%%%%%%%%%%%%%%%%%%%%%%%%%%%%%%%%%%%%%%%%%%%%%%%%%%
\begin{figure}[t]
\begin{center}
\includegraphics[angle=0,width= \textwidth]{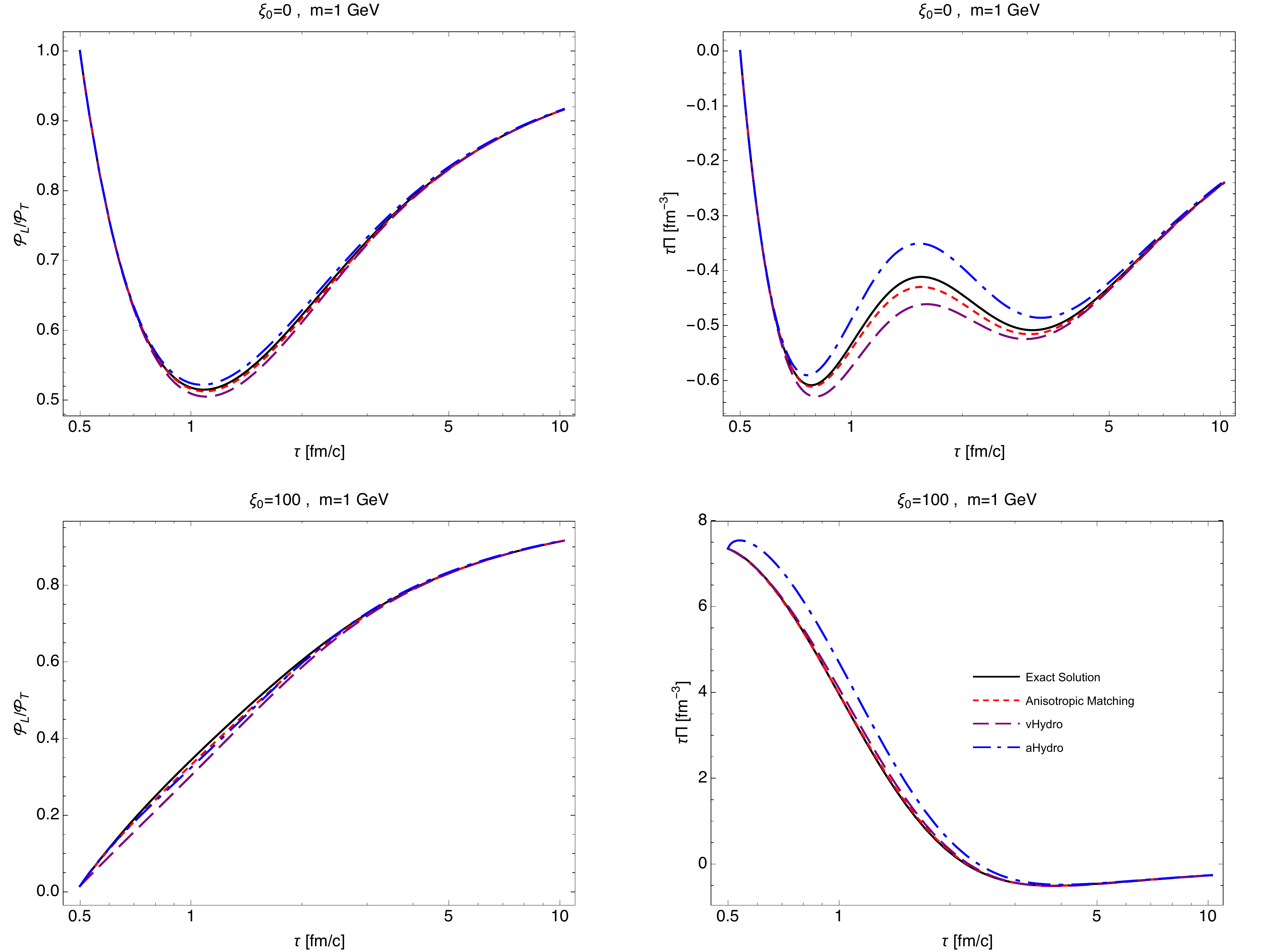}
\end{center}
\caption{(Color online) Same as Fig.~\ref{fig:03GeV}, except here we take $m = 1$ GeV.
}
\label{fig:1GeV}
\end{figure}
%%%%%%%%%%%%%%%%%%%%%%%%%%%%%%%%%%%%%%%%%%%%%%%%%%%%%%%%%%%%%%%%%%%%%%%%%%%%%%%%%%%%%%%%%%

We show the comparison with the exact solutions of three different approximations. By {\it anisotropic matching} we understand the leading order of the prescription presented in this paper, by {\it viscous hydrodynamics} we denote the modified gradient expansion by Jaiswal et al.~\cite{Jaiswal:2013npa} and, finally,  {\it anisotropic hydrodynamics} stands for the prescription introduced by Nopoush et al. in Ref.~\cite{Nopoush:2014nc}. The latter assumes the longitudinal boost invariance and cylindrically symmetric radial expansion. There is a recent, more general, formulation of anisotropic hydrodynamics~\cite{Tinti:2014yya} which does not require such symmetries in the evolution of the system. However, as it is shown in Ref.~\cite{Tinti:2015xra}, the less general prescription has a similar or even better agreement than the more general one in the $(0+1)$-dimensional framework.  On the other hand,  it is useful to remind that the viscous hydrodynamics prescription we choose, is the close-to-equilibrium limit of the equations seen in the last Section, therefore, we should expect a better agreement between the full equations and the exact solutions of the Boltzmann equation, than between the linearized version and the exact results. Confronting the last results found in the  literature (see, in particular, Refs~\cite{Denicol:2014sbc,Jaiswal:2014}), it is interesting to note that the prescription by Nopuosh et al. for anisotropic hydrodynamics and Jaiswal's prescription for the second order viscous hydrodynamics provide the best agreement with the exact solutions, excluding higher order treatment\footnote{For the next to leading order treatment of anisotropic  hydrodynamics see Refs.~\cite{Bazow:2013ifa,Bazow:2015cha}, and for viscous hydrodynamics see Ref.~\cite{Bazow:2013ifa,Chattopadhyay:2014lya}.}. In particular the anisotropic hydrodynamics prescription systematically reproduces the pressure anisotropy better, while the viscous hydrodynamics one  is more accurate for the isotropic pressure corrections.

Figures~\ref{fig:03GeV} and~\ref{fig:1GeV} clearly demonstrate that the approach providing the best approximation to the exact solution is the leading order of the anisotropic matching procedure, i.e.,  Eqs.~(\ref{e_ev}),~(\ref{U_ev}),~(\ref{shear_RTA_LO_1}) and~(\ref{bulk_RTA_LO_1}) applied to the $(0+1)$-dimensional expansion. The anisotropic matching provides a similar or better agreement for $\pres_L/\pres_T$ than anisotropic hydrodynamics, and at the same time a similar or better agreement for $\tau\Pi$ than viscous hydrodynamics. In particular, for large starting anisotropy, it removes the qualitatively wrong behavior of the bulk evolution at initial stages, which has been up to now a common feature of leading order anisotropic hydrodynamics, see for instance Ref.~\cite{Tinti:2015xra}. At last, as it is expected, the equation we propose reproduce the exact results better the viscous hydrodynamics prescription (which we remind correspond to the close to equilibrium limit of the anisotropic matching), especially for the pressure anisotropy. This suggests that the higher order corrections are properly taken into account using the new procedure already at the leading order.

A very interesting comparison can be made between Fig.~\ref{fig:1GeV} of the present paper and the plots having the same parameters in Ref.~\cite{Bazow:2015cha}\footnote{That is Fig.~1. Note, however, the log-log scale in $\pres_L/\pres_T$, and the different prescription for viscous hydrodynamics.}. The agreement with the exact solutions for the anisotropic expansion (still at  leading order in the present paper) is already very close to that found for the {\it viscous anisotropic hydrodynamics}, that is,  next to leading order anisotropic hydrodynamics, but without making use of the generalized Landau matching~(\ref{aLM_1}). It would be interesting to check in the future if this striking agreement with the exact solution is preserved in other dynamical situations, for instance, for different masses or different flow configurations ({\it e.g.} anisotropic hydrodynamics has already been tested for the Gubser flow~\cite{Nopoush:2015gf}).

%%%%%%%%%%%%%%%%%%%%%%%%%%%%%%%%%%%%%%%%%%%%%%%%%%%%%%%%%%%%%%%%%%%%%%%%%%%%%%%%%%%%%%%%%%%%%%%%%%%%%
%%%%%%%%%%%%%%%%%%%%%%%%%%%%%%%%%%%%%%%%%%%%%%%%%%%%%%%%%%%%%%%%%%%%%%%%%%%%%%%%%%%%%%%%%%%%%%%%%%%%%
\section{Summary and conclusions}
\label{sect:conc}
%%%%%%%%%%%%%%%%%%%%%%%%%%%%%%%%%%%%%%%%%%%%%%%%%%%%%%%%%%%%%%%%%%%%%%%%%%%%%%%%%%%%%%%%%%%%%%%%%%%%%
%%%%%%%%%%%%%%%%%%%%%%%%%%%%%%%%%%%%%%%%%%%%%%%%%%%%%%%%%%%%%%%%%%%%%%%%%%%%%%%%%%%%%%%%%%%%%%%%%%%%%

In this paper a novel method to treat the hydrodynamics expansion around an anisotropic background is introduced. Differently from previous works on this topic,  we propose an extended Landau matching for the anisotropic distribution. In other words, we assume that the anisotropic background is capable to reproduce the stress energy-momentum tensor, without the contribution of the higher order corrections~(\ref{aLM_0}). In the present work only the generalized Romatschke-Strickland form~(\ref{fa_1}) is considered as the leading order distribution, but our arguments can be easily adjusted to a different anisotropic distribution, if needed.

The use of the generalized matching allows us to take the dynamical equations of hydrodynamics directly from the exact time evolution of the moments of the Boltzmann {\it distribution} (not to confuse with the moments of the Boltzmann {\it equation}). Following this procedure it is necessary to include at least the evolution of the energy momentum tensor (second moment of the distribution), since Eq.~(\ref{Tmunu_ev}) includes local four-momentum conservation~(\ref{first_moment_0}). Even if the latter equations depend only on the stress-energy tensor itself (and, thus, on the leading order only, because of the generalized Landau matching), the other independent equations get some contribution from the full distribution function because of higher-orders moments. This is a common feature of the exact evolution of all moments of the distribution function, namely, some higher moments of the distribution always appear in their exact evolution equations~(\ref{exact_moments}). Therefore, the exact dynamics is given by an infinite set of coupled equations for the moments of the Boltzmann distribution. The solution of this puzzle is equivalent to the one used in viscous hydrodynamics, that is, one should make an approximation for the deviation from the anisotropic background having a finite number of degrees of freedom (which are connected directly with the moments of the full distribution itself). In this way one can, for instance, continue using the exact evolution of the moments of the distributions to close the system.

In the present paper we consider for practical applications only the simplest approximation, restricting ourselves to the leading order. Namely, the full distribution is assumed to be well reproduced by the anisotropic background without the need of higher order corrections. As a consequence, Eq.~(\ref{Tmunu_ev}) is enough to close the system of equations. We leave the selection of the most appropriate treatment of the next to leading order corrections to be addressed in further research.

We analyze in detail the case of the collisional term treated in relaxation-time approximation and prove that in the close-to-equilibrium the equations of motion reduce to the most accurate prescription for the second order viscous hydrodynamics. We find a striking agreement between the new approach and the exact solution of the Boltzmann equation in the Bjorken flow limit. Both the pressure anisotropy and the bulk evolution are significantly improved compared to the most successful approximation; namely Nopoush et al. version of anisotropic hydrodynamics for the pressure anisotropy, and Jaiswal modified Chapman-Enskog viscous hydrodynamics for the bulk ~\cite{Denicol:2014sbc,Jaiswal:2014}. These preliminary results suggest that it is necessary to  take higher orders approximations to find a better agreement with the exact solutions. For instance, next to leading order anisotropic hydrodynamics or third order viscous anisotropic hydrodynamics, as the ones in Refs.~\cite{Bazow:2013ifa,Bazow:2015cha}. In fact, comparing the plots having the same parameters in the present work and in Ref.~\cite{Bazow:2015cha}, the agreement with the exact solutions already seems very close to next to leading order anisotropic hydrodynamics.

%%%%%%%%%%%%%%%%%%%%%%%%%%%%%%%%%%%%%%%%%%%%%%%%%%%%%%%%%%%%%%%%%%%%%%%%%%%%%%%%%%%%%%%%%%%%%%%%%%%
\acknowledgments
%%%%%%%%%%%%%%%%%%%%%%%%%%%%%%%%%%%%%%%%%%%%%%%%%%%%%%%%%%%%%%%%%%%%%%%%%%%%%%%%%%%%%%%%%%%%%%%%%%%

I thank Mauricio Martinez, Rados\l aw Ryblewski, Wojciech Florkowski, Wojciech Broniowski, and Michael Strickland for clarifying discussions and interesting comments.  This work has been supported by Polish National Science Center grant No. DEC-2012/06/A/ST2/00390. 

%%%%%%%%%%%%%%%%%%%%%%%%%%%%%%%%%%%%%%%%%%%%%%%%%%%%%%%%%%%%%%%%%%%%%%%%%%%%%%%%%%%%%%%%%%%%%%%%%%%%%
%%%%%%%%%%%%%%%%%%%%%%%%%%%%%%%%%%%%%%%%%%%%%%%%%%%%%%%%%%%%%%%%%%%%%%%%%%%%%%%%%%%%%%%%%%%%%%%%%%%%%
\section{Appendix: Explicit formulas}
\label{sect:explicit}
%%%%%%%%%%%%%%%%%%%%%%%%%%%%%%%%%%%%%%%%%%%%%%%%%%%%%%%%%%%%%%%%%%%%%%%%%%%%%%%%%%%%%%%%%%%%%%%%%%%%%
%%%%%%%%%%%%%%%%%%%%%%%%%%%%%%%%%%%%%%%%%%%%%%%%%%%%%%%%%%%%%%%%%%%%%%%%%%%%%%%%%%%%%%%%%%%%%%%%%%%%%

Here we present the formulas used for the numeric calculations in the last section. Because of longitudinal boost invariance and homogeneity on the transverse plane, the four velocity reads for the Bjorken flow

\begin{equation}\label{U_0+1}
 U^{\mu}=(\cosh \eta,0,0,\sinh \eta), \qquad \qquad \eta = \frac{1}{2}\ln\!\left( \frac{t+z}{t-z} \right),
\end{equation}
being $\eta$ the space-time rapidity. All Lorentz scalars depend only on the longitudinal proper time

\begin{equation}
 \tau = \sqrt{ t^2 -z^2}.
\end{equation}
The scalar expansion and the shear stress therefore read

\begin{equation}\label{U_grad_0+1}
 \theta = \frac{1}{\tau}, \qquad \sigma^{\mu\nu} = \frac{1}{3\tau}\left( \frac{}{} X^\mu X^\nu + Y^\mu Y^\nu \right) -\frac{2}{3\tau} Z^\mu Z^\nu,
\end{equation}
where

\begin{equation}
 Z^\mu = (\sinh\eta, 0, 0, \cosh\eta),
\end{equation}
is the longitudinal direction, and $X^\mu$ and $Y^{\mu}$ are two arbitrary directions on the transverse plane, for instance the $x$ and $y$ direction in the lab frame. The vector fields $U^\mu$, $X^\mu$, $Y^\mu$ and $Z^\mu$ provide an orthonormal basis, which in the local rest frame reads

\begin{eqnarray}
 U^\mu = (1,0,0,0), \quad X^\mu = (0,1,0,0 ), \quad Y^\mu = (0,0,1,0), \quad Z^\mu = (0,0,0,1).
\end{eqnarray}
Because of the symmetry, even the distribution function~(\ref{fa_1}) is largely simplified in the $0+1$-dimensional setup. Indeed, seen from the local rest frame (LRF)

\begin{equation}
 f_a = k \exp\left[ -\frac{1}{\lambda}\sqrt{ m^2\left( \frac{}{} 1+\phi \right) + \left( \frac{}{} 1-\frac{1}{2}\xi \right)p_T^2 + \left( \frac{}{} 1+\xi \right)p_L^2 } \right],
\end{equation}
being $p_L = (p\cdot Z)$ the longitudinal momentum, and $p_T$ the transverse momentum. It is convenient at this point to rewrite the last formula in the following way

\begin{equation}\label{fa_0+1}
 f_a = k \exp\left[ -\frac{1}{\Lambda}\sqrt{m^2 -\frac{p_T^2}{\alpha_T^2} + \frac{p_L^2}{\alpha_L^2} } \right], 
\end{equation}
where

\begin{equation}
\frac{1}{\Lambda^2} = \frac{1+\phi}{\lambda^2}, \qquad \frac{1}{\alpha_T^2} = \frac{1-\frac{1}{2}\xi}{(1+\phi)}, \qquad \frac{1}{\alpha_L^2} = \frac{1+\xi}{(1+\phi)}.
\end{equation}
Just like for the shear stress $\sigma^{\mu\nu}$, the symmetry of the expansion constrains $\pi^{\mu\nu}$ to depend on a single scalar function $\pi_s$

\begin{equation}\label{pi_0+1}
 \pi^{\mu\nu} = \frac{1}{2}\pi_s(\tau)\left( \frac{}{}X^\mu X^\nu +Y^\mu Y^\nu \right) -\pi_s(\tau) Z^\mu Z^\nu.
\end{equation}
It is convenient to introduce the the longitudinal pressure $\pres_L$ and the transverse pressure $\pres_T$, 
\begin{equation}\label{PLT}
 \pres_L = Z\cdot T\cdot Z = \preseq +\Pi -\pi_s, \qquad \qquad \pres_T = X\cdot T \cdot X = Y\cdot T\cdot Y = \preseq + \Pi + \frac{1}{2}\pi_s,
\end{equation}
since this results in relatively simple expressions, which do not depend on the effective temperature, namely

\begin{eqnarray}\label{PL}
 \pres_L &=& k \int d^3{\bf p} \, \frac{p_L^2}{ \sqrt{m^2 +{\bf p}^2}}\,  \exp\left( -\frac{1}{\Lambda} \sqrt{m^2 + \frac{p_T^2}{\alpha_T^2} + \frac{p_L^2}{\alpha_L^2}} \right) = {\cal \tilde{H}}_{3L}\Lambda^4, \\\label{PT}
 \pres_T &=& \frac{k}{2} \int d^3{\bf p} \, \frac{p_T^2}{ \sqrt{m^2 +{\bf p}^2}}\,  \exp\left( -\frac{1}{\lambda} \sqrt{m^2 + \frac{p_T^2}{\alpha_T^2} + \frac{p_L^2}{\alpha_L^2}} \right) = {\cal \tilde{H}}_{3T}\Lambda^4.
\end{eqnarray}
Here the integrals are taken in the local rest frame. Having the same anisotropic distribution for the same (subset of) equations, the energy and momentum conservation reduce to the single equation originally found in Ref.~\cite{Nopoush:2014nc}, namely

\begin{equation}
 D\ped = -\frac{\ped +\pres_L}{\tau},
\end{equation}
or, writing explicitly the derivatives on the variables $\Lambda$, $\alpha_T$ and $\alpha_L$
\begin{equation}
 \left( \frac{}{} 4{\cal \tilde{H}}_3 -\tilde{\Omega}_m \right)D\ln\Lambda + 2\,  \tilde{\Omega}_T \, D\ln\alpha_T +\tilde{\Omega}_L \,\left( \frac{1}{\tau} + D\ln\alpha_L \right) =0.
\end{equation}
The explicit expressions of the functions $\tilde{\cal H}_{3}$, $\tilde{\cal H}_{3T}$, $\tilde{\cal H}_{3L}$, $\tilde{\Omega}_m$, $\tilde{\Omega}_T$, and $\tilde{\Omega}_L$ are the ones found in Ref.~\cite{Nopoush:2014nc} itself. They are all derived from the exact formula for the proper energy density in $0+1$-dimensions

\begin{equation}
 \ped = \Lambda^4\tilde{\cal H}_{3} = (2\pi \, k) \, \alpha_T^4 \, \Lambda^4 \int_0^{\infty} d r \, r^3\exp\left[ -\sqrt{ {\hat m}^2 + r^2} \right] {\cal H}_{2}(y, z),
\end{equation}
being

\begin{equation}
 {\hat m} = \frac{m}{\Lambda}, \qquad y= \frac{\alpha_L}{\alpha_T},\qquad z=\frac{{\hat m}}{\alpha_T r}, 
\end{equation}
and

\begin{eqnarray}\nonumber
 {\cal H}_2 (y,z) &=& y \int_{-1}^1 d(-\cos \theta)\sqrt{z^2 +y^2\cos^2\theta + \sin^2 \theta} = \\ \label{H2}
 && \qquad= \frac{y}{\sqrt{y^2-1}}\left[ (1+z^2) \tanh^{-1}\sqrt{\frac{y^2-1}{y^2+z^2}} + \sqrt{(y^2 +z^2)(y^2 -1)}\right].
\end{eqnarray}
The other other functions are

\begin{eqnarray}
 \tilde{\cal H}_{3L} &=& (2\pi \, k) \alpha_T^4 \int_0^\infty dr  \, r^3\exp\left[ -\sqrt{{\hat m}^2 + r^2} \right] {\cal H}_{2_L}(y, z), \\
 \tilde{\cal H}_{3T} &=& \frac{1}{2}(2\pi \, k) \alpha_T^4 \int_0^\infty dr  \, r^3\exp\left[ -\sqrt{ {\hat m}^2 + r^2} \right] {\cal H}_{2_T}(y, z), \\
 \tilde{\cal H}_{3m} &=& (2\pi \, k) \alpha_T^4 \, {\hat m}^2\int_0^\infty dr  \frac{ r^3}{\sqrt{{\hat m}^2 + r^2} }\exp\left[ -\sqrt{ {\hat m}^2 + r^2} \right] {\cal H}_{2}(y, z),
\end{eqnarray}
while

\begin{eqnarray}
 {\cal H}_{2L}(y,z) &=& y\partial_y{\cal H}_2(y,z) - {\cal H}_2(y,z), \\
 {\cal H}_{2T}(y,z) &=& {\cal H}_2(y,z) -{\cal H}_{2L}(y,z) -z\partial_z{\cal H}_2(y,z) ,
\end{eqnarray}
and

\begin{eqnarray}
 \frac{\partial \tilde{\cal H}_3}{\partial\alpha_T} &=& \frac{2}{\alpha_T}\left[ \frac{}{} \tilde{\cal H}_3 + \tilde{\cal H}_{3T} \right] \equiv \frac{2}{\alpha_T}\tilde{ \Omega}_T, \\
 \frac{\partial \tilde{\cal H}_3}{\partial\alpha_T} &=& \frac{1}{\alpha_L}\left[ \frac{}{} \tilde{\cal H}_3 + \tilde{\cal H}_{3L} \right]\equiv \frac{1}{\alpha_L}\tilde{\Omega}_L, \\
 \frac{\partial \tilde{\cal H}_3}{\partial\alpha_T} &=& \frac{1}{{\hat m}}\left[ \frac{}{} \tilde{\cal H}_3 - \tilde{\cal H}_{3L}- 2 \tilde{\cal H}_{3T} - \tilde{\cal H}_{3m} \right]\equiv\frac{1}{\hat m} \tilde{\Omega}_m.
\end{eqnarray}
Taking into account the symmetry of~(\ref{fa_0+1}) and making use of~(\ref{U_grad_0+1}),~(\ref{pi_0+1}) and~(\ref{PLT}), we can rewrite equations~(\ref{shear_RTA_LO_1}) and~(\ref{bulk_RTA_LO_1}) in the following way

\begin{eqnarray} \label{T-L_0}
 D\left( \pres_T -\pres_L \right) &=& \frac{1}{\tau} \Lambda^4\left( \frac{}{} \tilde{\cal H}_{3TL} - \tilde{\cal H}_{3LL} \right) -\left( \frac{1}{\tau} +\frac{1}{\taueq} \right)\left( \frac{}{} \pres_T - \pres_L \right) +\frac{2}{\tau} \pres_L, \\ \label{2T+L_0}
 D\left( \frac{}{} 2\pres_T -\pres_L \right) &=& \frac{1}{\tau}\Lambda^4\left( \frac{}{} 2\tilde{\cal H}_{3TL} +  \tilde{\cal H}_{3LL} \right) -\left( \frac{1}{\tau} +\frac{1}{\taueq} \right) \left( \frac{}{} 2\pres_T +\pres_L \right) -\frac{2}{\tau}\pres_L +\frac{3}{\taueq}\preseq.
\end{eqnarray}
The functions $\tilde{\cal H}_{3TL}$ and $\tilde{\cal H}_{3LL}$ are defined by

\begin{eqnarray}
 \tilde{\cal H}_{3LL} &=& \frac{1}{\Lambda^4}\int dP \frac{(p\cdot Z)^4}{(p\cdot U)^2}f_a \equiv (2\pi \, k) \, \alpha_T^4 \, \int_0^\infty dr \, r^3 \exp\left[ -\sqrt{ {\hat m}^2 + r^2} \right] \tilde{\cal H}_{2LL}, \\
 \tilde{\cal H}_{3TL} &=& \frac{1}{\Lambda^4}\int dP \frac{(p\cdot X)^2(p\cdot Z)^2}{(p\cdot U)^2}f_a \equiv \frac{1}{2}(2\pi \, k) \, \alpha_T^4 \, \int_0^\infty dr \, r^3 \exp\left[ -\sqrt{ {\hat m}^2 + r^2} \right] \tilde{\cal H}_{2TL},
\end{eqnarray}
being

\begin{equation}
 {\cal H}_{2TL} = {\cal H}_{2T} -y\partial_y{\cal H}_{2T}, \qquad {\cal H}_{2LL} = 3 {\cal H}_{2L} -y\partial_y{\cal H}_{2L}.
\end{equation}
In order to solve Eqs.~(\ref{T-L_0}) and~(\ref{2T+L_0}) it is necessary to have an explicit formula for the derivatives of the longitudinal and transverse pressure

\begin{eqnarray}
 D\pres_L &=& \Lambda^4 \left[ \left( \frac{}{} 4\tilde{\cal H}_{3L} -\tilde{\Omega}_{Lm} \right) D\ln \Lambda + \tilde{\Omega}_{LT} D\ln \alpha_T + \tilde{\Omega}_{LL}D \ln \alpha_L \right], \\
 D\pres_T &=& \Lambda^4 \left[ \left( \frac{}{} 4\tilde{\cal H}_{3T} -\tilde{\Omega}_{Tm} \right) D\ln \Lambda + \tilde{\Omega}_{TT} D\ln \alpha_T + \tilde{\Omega}_{TL}D \ln \alpha_L \right],
\end{eqnarray}
being

\begin{eqnarray}
 \tilde{\Omega}_{Lm} = {\hat m} \, \partial_{\hat m} \,  \tilde{\cal H}_{3_L} , \qquad &  \tilde{\Omega}_{LT} = {\alpha_T} \, \partial_{\alpha_T} \,  \tilde{\cal H}_{3_L} , & \qquad \tilde{\Omega}_{LL} = {\alpha_L} \, \partial_{\alpha_L} \,  \tilde{\cal H}_{3_L} , \\
 \tilde{\Omega}_{Tm} = {\hat m} \, \partial_{\hat m} \,  \tilde{\cal H}_{3_T} , \qquad &  \tilde{\Omega}_{TT} = {\alpha_T} \, \partial_{\alpha_T} \,  \tilde{\cal H}_{3_T} , & \qquad \tilde{\Omega}_{TL} = {\alpha_L} \, \partial_{\alpha_L} \,  \tilde{\cal H}_{3_T}.
\end{eqnarray}
Introducing the quantities

\begin{eqnarray}
 && \tilde{\cal H}_{3Lm} = (2\pi \, k) \,\alpha_T^4 \int dP \, \frac{{\hat m}^2}{\sqrt{{\hat m}^2 + r^2}} \exp\left[ -\sqrt{{\hat m}^2 + r^2} \right] {\cal H}_{2L}, \\
 && \tilde{\cal H}_{3Tm} =\frac{1}{2} (2\pi \, k) \,\alpha_T^4 \int dP \, \frac{{\hat m}^2}{\sqrt{{\hat m}^2 + r^2}} \exp\left[ -\sqrt{{\hat m}^2 + r^2} \right] {\cal H}_{2T}, \\
 && \tilde{\cal H}_{3TT} =\frac{1}{4} (2\pi \, k) \,\alpha_T^4 \int dP \, \exp\left[ -\sqrt{{\hat m}^2 + r^2} \right] {\cal H}_{2TT},
\end{eqnarray}
with

\begin{equation}
 {\cal H}_{2TT} = y\, \partial_y \, {\cal H}_{2T} +z\, \partial_z \, {\cal H}_{2T},
\end{equation}
it is possible to write all of the $\tilde{\Omega}$'s functions as integrals of the derivatives of~(\ref{H2}).

\begin{eqnarray}
 && \tilde{\Omega}_{Lm} = \tilde{\cal H}_{3LL} + 2 \tilde{\cal H}_{3TL} - \tilde{\cal H}_{3L} - \tilde{\cal H}_{3Lm}, \\
 && \tilde{\Omega}_{LL} = 3\tilde{\cal H}_{3L} - \tilde{\cal H}_{3LL}, \\
 && \tilde{\Omega}_{LT} =  2 \tilde{\cal H}_{3L} - 2\tilde{\cal H}_{3TL}, \\
 && \tilde{\Omega}_{Tm} = 2\tilde{\cal H}_{3TT} +  \tilde{\cal H}_{3TL} - \tilde{\cal H}_{3T} - \tilde{\cal H}_{3Tm}, \\
 && \tilde{\Omega}_{TL} = \tilde{\cal H}_{3T} - \tilde{\cal H}_{3TL}, \\
 && \tilde{\Omega}_{TT} = 4\tilde{\cal H}_{3T} - 2 \tilde{\cal H}_{3TT}. \\
\end{eqnarray}

\end{document}